\documentclass[aps,prl,twocolumn,superscriptaddress,showpacs,amsmath,amssymb,longbibliography]{revtex4-1}
\usepackage{graphicx}
\usepackage[breaklinks]{hyperref}
\usepackage[english]{babel}
\usepackage{amsmath}
\usepackage{xcolor}
\usepackage{ulem}

\usepackage{color}
\definecolor{nred} {RGB}{224,0,0}
\definecolor{nblue} {RGB}{28,130,185}
\definecolor{dgreen} {RGB}{78,138,21}

\usepackage[sort&compress]{natbib}
\begin{document}


\title{Einstein relation  for a driven disordered  quantum chain in  subdiffusive regime }

\author{M. Mierzejewski}
\affiliation{Department of Theoretical Physics, Faculty of Fundamental Problems of Technology, Wroc\l aw University of Science and Technology, 
50-370 Wroc\l aw, Poland}

\author{P. Prelov\v sek}
\affiliation{J. Stefan Institute, SI-1000 Ljubljana, Slovenia }
\affiliation{Faculty of Mathematics and Physics, University of Ljubljana, SI-1000 Ljubljana, Slovenia }
\author{J. Bon\v ca}
\affiliation{Faculty of Mathematics and Physics, University of Ljubljana, SI-1000 Ljubljana, Slovenia }
\affiliation{J. Stefan Institute, SI-1000 Ljubljana, Slovenia }

\begin{abstract}
A quantum particle propagates
subdiffusively  on a strongly disordered chain when it is coupled to itinerant hard-core bosons. 
We establish a  generalized Einstein relation (GER) that  relates such subdiffusive spread
to an unusual  time-dependent drift velocity,  which appears as a consequence of   a constant electric field. We show that  GER remains valid much beyond the regime of the linear response.
Qualitatively, it holds true up to strongest drivings when the nonlinear field--effects lead to the Stark-like localization.
Numerical calculations based on full quantum evolution are  substantiated  by much simpler rate equations for the 
boson-assisted transitions between localized  Anderson states.


\end{abstract}

\maketitle

{\it Introduction--}
Over  two  decades after the outstanding  discovery of the Anderson localization (AL) phenomena \cite{anderson58}, the effect of the interplay between the disorder and  many-body interactions on transport properties of metals started to be recognized as one of the fundamental unsolved problems in solid state physics \cite{ramakrishan1985,fleishman80}.  The importance of interactions on AL systems is now identified  as a concept   of  the many-body localization (MBL) \cite{basko06,oganesyan07}.  The presence of MBL  has been theoretically confirmed predominantly  in systems that posses only spin or charge degrees of freedom   
\cite{monthus10,luitz15,ZZZ5_4,Ponte2015,lazarides15,vasseur15a,serbyn2014a,pekker2014,torres15,torres16,laumann2015,huse14,gopal17,Hauschild_2016,herbrych13,imbrie16,steinigeweg16}. Moreover, the existence of MBL   has been found  in a few experimental 
studies \cite{kondov15,schreiber15,choi16,bordia16,bordia2017_1,smith2016}.  Among several characteristic features of strongly disordered systems  is unusually slow  time evolution of characteristic physical properties \cite{znidaric08,bardarson12,kjall14,serbyn15,luitz16,serbyn13_1,bera15,altman15,agarwal15,gopal15,znidaric16,mierzejewski2016,lev14,lev15,barisic16,bonca17,bordia2017_1,zakrzewski16,protopopov2018,sankar2018,zakrzewski2018}  that typically   emerges as  subdiffusive dynamics as a precursor to MBL transition \cite{luitz2016prl,luitz116,znidaric16,gopal17,kozarzewski18,prelovsek217,new_karrasch,prelovsek2018a}. 

 In this Letter we consider the effect of driving (via the constant electric field) on a quantum particle in a random chain 
coupled to itinerant hard-core bosons (HCB).   We note that such a system simulates the propagation of a (single) charge  
coupled to spin degrees in strongly correlated  systems, as e.g., the disordered Hubbard type-models 
\cite{mondaini15,prelovsek16,bonca17},
being realized in cold-atom experiments \cite{kondov15,schreiber15,choi16,bordia16,bordia2017_1,smith2016}.
 It has long been  assumed that  the AL phenomenon is destroyed by the electron--phonon coupling via the mechanism of 
 phonon-assisted hopping \cite{mott1968, emin1975}.  Recently the absence of localization and the onset of normal diffusion of a particle coupled to standard itinerant bosons has been confirmed via  a direct quantum evolution \cite{sante2017}. Still,  the itinerant HCB appear to be a separate  case with a transient  or even persistent  subdiffusive dynamics \cite{bonca2018,prelovsek2018a}. 
 While the subdiffusive dynamics has been found in various disordered interacting systems, the behavior of such  system under constant driving 
 remains predominantly unexplored \cite{kozarzewski16}  whereby  in driven MBL systems the focus has been mostly
 on periodic drivings \cite{Ponte2015,Ponte2015a,Bordia2017,Agarwal2017,Shen2017}. 

Transport properties of disordered quantum interacting many-body systems  depend on  the disorder strength. 
Weakly disordered systems typically display generic transport properties. In particular,   a one-dimensional (1D)  chain reveals 
normal diffusion  \cite{znidaric16}, i.e.,  a nonuniform particle density spreads as $\langle x^2(t) \rangle_0 = 2D  t$, where $D$ 
is the diffusion constant. A weak external field, $F$, induces  a drift,  $\langle x(t) \rangle_F = \mu F t$, where 
$\mu$ is the mobility. According to the  Einstein relation \cite{zwanzig}, $\mu= \beta D$,  
where $\beta=1/{k_B T}$ with temperature $T$,
the relation being valid also for quantum particles at high-enough $T$.  

Strongly disordered chains of spinless fermions show  MBL  when the  d.c. transport is completely suppressed.
Here, we are interested in the intermediate case when
the particles spread subdiffusively, i.e., with the d.c. value $D_0=0$ but $\langle x^2(t) \rangle_0 \propto t^{\gamma}$
with $0<\gamma <1$. There is a vast theoretical evidence for  such subdiffusive evolution without external driving 
in disordered 1D systems, whereby the anomalous spread has been explained via the "weak--link" 
scenario \cite{agarwal15,bordia2017_1,agarwal16,luschen17}. 
Despite $D_0=0$, one expects the relevance of generalized Einstein relation (GER) \cite{robin2009,bouchaud89}
\begin{equation}
\langle x(t) \rangle_F= F \frac{\beta}{2} \langle x^2(t) \rangle_0, \label{ger}
\end{equation}       
as long as the system remains in the linear response (LR) regime.  However,  subdiffusive systems display very slow relaxation, 
hence even a weak field can drive the system far from equilibrium. Then, $\beta$ in Eq.~(\ref{ger}) might be ill defined,
hence  the limits of the LR regime and the applicability of Eq.~(\ref{ger}) should be methodically  explored.  

In the following we  show that the GER holds true even for large fields within the quasiequlibrium LR when   the  temperature increases in time due to heating. For  even stronger fields the particle dynamics gradually  slows down (approaching the effective exponent $\gamma=0$), related in this regime to the phenomenon of  Stark localization.  
 We also show that results obtained via full quantum evolution can be well explained with much simpler rate equations, 
where the transition rates between Anderson states are evaluated via the Fermi golden rule (FGR).   

{\it Model and method--}
 We study a model of  quantum particle moving in a disordered chain (Anderson model) 
and  coupled to bosons,
\begin{eqnarray}
H &=& - t_h \sum_j (c^\dagger_{j+1}  c_j + {\mathrm h.c.}) + \sum_j h_j n_j  + g \sum_j n_j 
( a^\dagger_j + a_j ) \nonumber \\
&+& \omega_0 \sum_j a_j^\dagger a_j - t_b  \sum_j ( a_{j+1}^\dagger a_j + \mathrm{H.c.}).
\label{ham} 
\end{eqnarray}   
where $n_j=c^\dagger_j c_j$ is local particle density and the random potential $h_j$  is assumed to be uniformly distributed in $[-W,W]$.  
Bosons in Eq.  (\ref{ham}) are itinerant   due to finite hopping $0<t_b < \omega_0/2$.
We consider (predominantly)  bosons  being HCB with only two states per site.
However,  we briefly discuss also the coupling to standard bosons which  leads to a normal diffusive transport \cite{prelovsek2018a}. 

In order to study the particle driven with a constant electric field $F$ one  considers either   a system with periodic boundary 
conditions (p.b.c.) and  time--dependent Peierls phase $t_h \rightarrow  t_h \exp(-i F t)$  or
a chain with open  boundary conditions  and additional electrostatic potential $H \rightarrow H -\sum_j  j  F  n_j$. 
While both approaches are equivalent \cite{jim2005},  the former one is more convenient for full quantum dynamics and the 
latter one with time--independent $H$ facilitates calculations of  the transition rates between the Anderson states.  
Finally,  we take $|t_h|=1$ as the energy unit.   

{\it Numerical results --}
First, we numerically study the Hamiltonian, Eq.~(\ref{ham}), with a Peierls driving $t_h(t)= \exp(-i F t)$, $F=$const. 
We consider only relatively strong disorder, $W  \ge 3 $. It has been shown in Ref. \cite{prelovsek2018a} that in this regime 
(at $F=0$) the particle coupled to HCB spreads subdiffusively, i.e.,  $ \langle x^2(t) \rangle_0 \propto t^\gamma$ with $\gamma <1$.   
The key question is whether the GER, Eq.~(\ref{ger}), holds true and the transport anomaly shows up in the current that is 
induced by $F \neq 0$.  In principle, Eq. (\ref{ger}) can be directly tested only for $F \rightarrow 0$, when $\beta(t) =\beta (0)$.
otherwise the system's energy  increases due to driving,  $\Delta E(t)=F \langle x(t) \rangle_F$.
Still, recalling that in the high--temperature regime $T \gg t_h$ the kinetic energy is proportional to the inverse temperature 
$E_k(t) \propto -\beta(t) $, where $E_k=\langle   \sum_j e^{-i F t} c^\dagger_{j+1}  c_j \rangle + {\mathrm c.c.} $,  the instantaneous $\beta(t)$
can be also directly monitored.
Utilizing the latter proportionality,  we rewrite Eq. (\ref{ger}) in a form that may be directly tested also for driven closed systems
\begin{equation}
R(t)=-\frac{ \Delta  E(t)}{F^2  E_k(t)} \propto \langle x^2(t) \rangle_0  \propto t^{\gamma}.
\label{ratio}
\end{equation}

\begin{figure} 
 \includegraphics[width=\columnwidth]{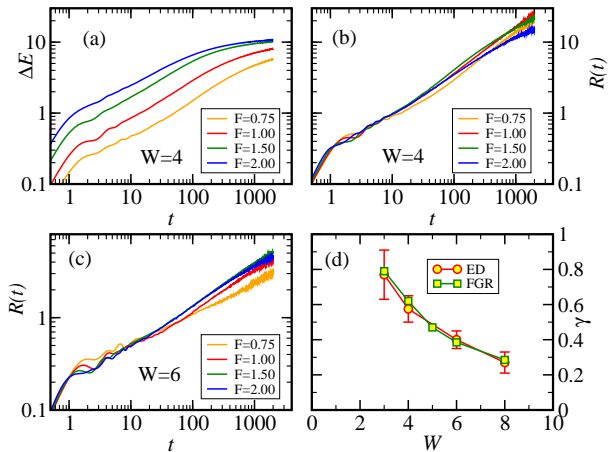}
\caption{ Results  of full time propagation on  systems with $L=14$  for energy increase  $\Delta E(t)$ in (a), 
and $R(t)$ in (b) and (c).  In (d)  $\gamma$  vs. $W$ is shown, as  obtained from fitting $R(t)\propto t^\gamma$
for $t\gtrsim 1$ in case of ED (circles) and using FGR (squares). In the latter case $\gamma$  was extracted from 
$\langle x^2(t) \rangle_{0}\propto t^\gamma$, see Ref.~\cite{prelovsek2018a}. Note that unless otherwise specified, we have used: $\omega_0=g=1$, and $t_g=0.5$.}   
\label{fig1}
\end{figure}

Numerical calculations were performed on 1D systems  with up to  $L=14$ sites  with p.b.c. The  size of the Hilbert space is given
 by $N_\mathrm{st}=L~2^L$, whereas the finite--size effects are discussed in the Supplemental Material \cite{supp}. Both energies in  Eq.~(\ref{ratio}) were obtained by sampling over $N_\mathrm{s}=10^3$ realizations 
of disorder. Full quantum time evolutions were performed while taking the advantage of the Lanczos technique \cite{park}  starting from 
the corresponding ground state of $H$. To achieve sufficient accuracy of time propagation, we used time step $\Delta t = 0.02$ and 
reached times $t \sim 10^3$.   In Fig.~\ref{fig1}(a) we show the sample-averaged energy increase
$\Delta E(t) = (1/N_\mathrm{s})\sum_i (E_i(t)- E_i(0))$, 
where $E_i(0)$ is the  ground-state energy of $H$, Eq.~(\ref{ham}), corresponding to $i$-th particular random-potential configuration 
$\{h_j \}$. Figs. \ref{fig1}(b) and (c) show the ratio $R(t)$ for two values of disorder $W=4, 6$. The main observation is that results are consistent 
with $R(t) \propto t^\gamma$, based on the assumption of the GER. In  Fig. ~\ref{fig1} (d) we show   extracted exponents $\gamma$ for different values of $W$ along with those obtained from the spread of  $\langle x^2(t) \rangle_0$  using  analytical approach based on FGR, as discussed below. 

\begin{figure} 
 \includegraphics[width=\columnwidth]{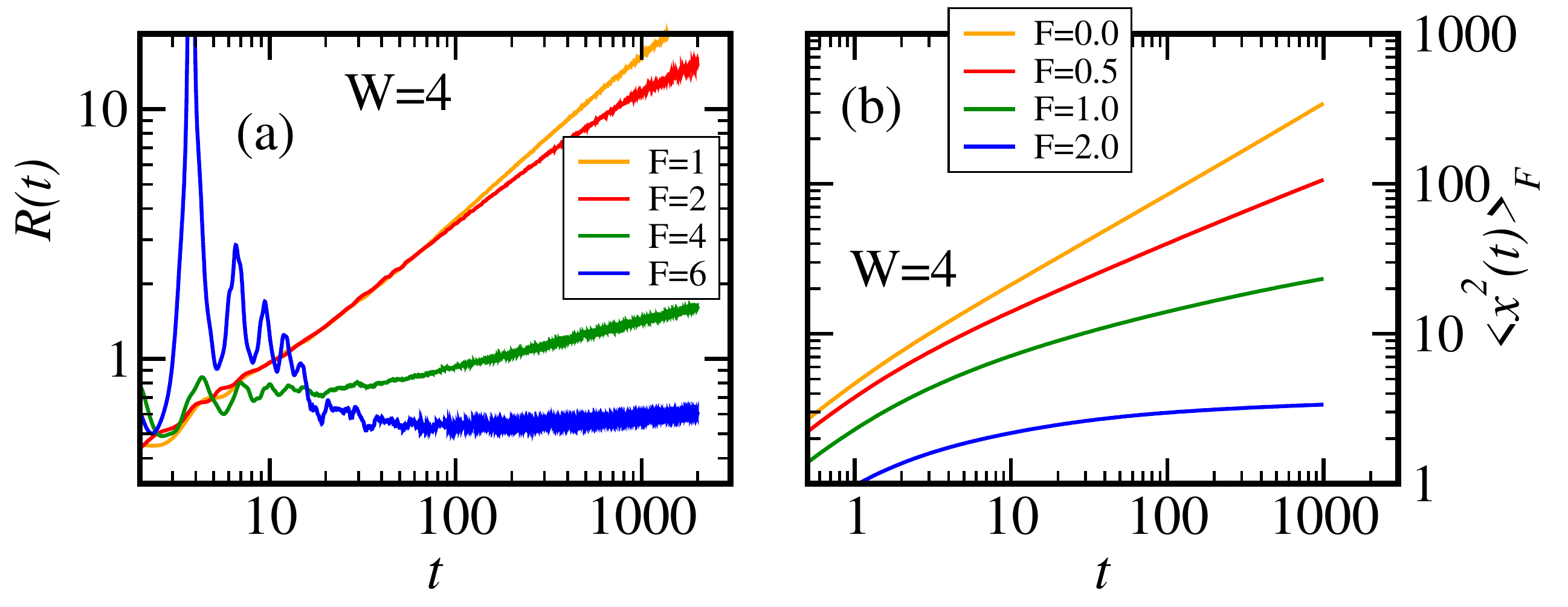}
\caption{ In (a) we show $R(t)$ for $W=4$ for different $F > 0$, up to maximal $F=6$ using ED on $L=14$ sites,
and in (b) the spread $\langle x^2(t) \rangle_{F}$,    using FGR for $L=400$. }   
\label{fig2}
\end{figure} 

Fig. \ref{fig2}a shows numerical results for $R(t)$ for a fixed disorder strength $W$ but various drivings $F$. 
It is evident that for stronger driving $F>2$, exponent $\gamma$ (as well as particle dynamics) is significantly reduced
and eventually, for very strong $F \sim 6 $ the particle becomes almost localized, here due to the Stark phenomenon.  
%

{\it Dynamics via rate equations.--}
In the following we describe the system's dynamics using the rate equations for the transitions between the localized 
Anderson states.  The  method has been shown to reproduce for $F=0$ the subdiffusive particle dynamics as well as
(at least qualitatively) values for the exponent  $\gamma$, \cite{prelovsek2018a}.  Direct comparison of numerical results obtained
from  ED and  the rate equations is shown in the Supplemental Metarial \cite{supp}.
 We stress that it is essential to study the distribution of the 
transition rates and not just their average values, since the averaging erases the essential information on their large and singular fluctuations.
In particular,  averaging of localized and  delocalized samples would (mistakenly) indicate that the particle is always delocalized.  

Here, we recall only main steps of Ref. ~\cite{prelovsek2018a}  for derivation of the transition rates,  now taking into account $F \neq 0$.
First we solve  the single--particle eigenproblem  for open boundary conditions
 \begin{equation}
 \sum_j \left[ -t_h(c^\dagger_{j+1}  c_j+ {\mathrm h.c.}) +  (h_j - j F) n_j \right]=\sum_{l} \epsilon_l \varphi^{\dagger}_l  \varphi_l,
 \end{equation}
where $ \varphi^{\dagger}_l = \sum_j \phi_{lj} c^{\dagger}_j$ creates a particle in the localized  Anderson--Stark  
state $| l \rangle$.  Using the FGR one obtains transition rates between states $l, l'$ which originate from the coupling to HCB,
\begin{eqnarray}
\Gamma_{l,l'} 
&=& \int_{-\infty}^{\infty} {\mathrm d} t \; e^{-it\Delta_{l,l'}} \frac{1}{L} \sum_q |\eta_{l,l',q} |^2 \nonumber \\
&&  \times \left[ f(\omega_q) e^{-i\omega_q t} +f(-\omega_q) e^{i\omega_q t}   \right],  \label{gam1} \\
\eta_{l,l',q} &= & g \sum_{j}  \mathrm{e}^{-iqj}    \phi_{l'j}  \phi_{lj}, \label{gam2}
\end{eqnarray}
where $\omega_q=\omega_0-2t_B \cos(q)$,  $f(\omega_q)$ is the Fermi-Dirac distribution function and  $\Delta_{l,l'}=\epsilon_{l}-\epsilon_{l'}$.  
Within the FGR, the particle dynamics can be described via the rate equations for occupations  
$n_l(t) =\langle  \varphi^{\dagger}_l  \varphi_l \rangle(t) $,
\begin{equation}
\frac{{\rm d} n_l}{{\rm d} t} = \sum_{l'} ( \Gamma_{l'l} n_{l'} - \Gamma_{ll'} n_l). \label{req}   
\end{equation}   

Fig. \ref{fig2}b shows the  evolution of the averaged spread, $\langle x^2(t) \rangle_{F} =\sum_l (x_l-x_{l_0})^2 n_l(t)$, 
where $x_l=\sum_j j |\phi_{lj}|^2$, as obtained  via rate equations at $\beta \to 0$ for $L=400$ and open boundary conditions. 
Particle is  initially  put in the middle of the system, well away from the
boundaries, i.e., $n_l(0)=\delta_{l,l_0} $. 
Due to strong disorder, $W=4$, the particle spread is subdiffusive already for  $F=0$. For larger $F > 1$, the diffusion further slows down 
and eventually for $F \geq 2$ the particle tends to localize due to the Stark effect, i.e., $\langle x^2(t \to \infty) \rangle_{F}$  saturates. 
We stress a clear qualitative similarity between  the spread  and the rescaled drift,   $R(t)$,  shown in Fig. \ref{fig2}a. 
This similarity persists even when both quantities  are determined for strong $F$,  well beyond the the LR regime relevant for Eq. (\ref{ger}).  
  
Eqs.~(\ref{gam1})-(\ref{req})   can be studied numerically even for large systems. However, in order to derive the GER  within rate-equation approach, 
we rewrite Eq.~(\ref{gam1}) in a more symmetric form, representing it  as $\Gamma_{l,l'}=\Gamma^{0}_{l,l'} [1+\tanh(\beta \Delta_{l,l'}/2)] $, where $\Gamma^{0}_{l,l'}$ refer to rates for $\beta \rightarrow 0$,
 \begin{equation}
 \Gamma^{0}_{l,l'}=\frac{\pi}{L} \sum_q  |\eta_{l,l',q} |^2  \left[ \delta(\omega_q+\Delta_{l,l'})+ \delta(\omega_q -\Delta_{l,l'}) \right].
 \label{gammafull}
 \end{equation}
 In general,  $F \neq 0$  enters  via the (symmetric) overlaps, $\eta_{l,l',q}=\eta_{l',l,q}$, and the (antisymmetric) 
 energy differences, $\Delta_{l,l'}=-\Delta_{l',l}$.  The antisymmetry of $\Gamma_{l,l'}$ originates solely from $\tanh(\beta \Delta_{l,l'}/2)$.
The time-evolution of the particle drift due to $F\neq 0$ can be evaluated as the current, 
$ I(t)= {\rm d }\langle x(t) \rangle/{\rm dt} =\sum_l x_l  {\rm d }n_l / {\rm dt}$.  Using Eq.~(\ref{req}) one then obtains
\begin{eqnarray}
I(t) &= & - \sum_{l,l'}   \frac{  (x_l-x_{l'})^2}{2}  \Gamma^0_{l,l'} \times \nonumber \\  
&& \left[ \frac{n_l-n_{l'}}{x_l-x_{l'}}  +  (n_l+n_{l'})    \frac{\tanh[\frac{ \beta (\epsilon_l-\epsilon_{l'})}{2}]}{x_l-x_{l'}} \right] .
\label{einstein}
\end{eqnarray}
For high $T$, the second term may be expanded in $\beta$. Then,  Eq. ~(\ref{einstein}) becomes the Einstein relation 
which  states that the current is  induced by the gradient of the particle density  (first term)  and  the gradient of the
potential (second term),  where the latter is weighted by $\beta$ and the local density, $(n_l+n_{l'})/2$.

For strong disorder and weak $F$ one may assume that $\epsilon_l \simeq \epsilon^0_l-x_l F $,
where $\epsilon^0_l$ refers the Anderson state at $F=0$. Since  on average $\epsilon^{0}_l-\epsilon^{0}_{l'}$ vanishes,
the term $(\epsilon^{0}_l-\epsilon^{0}_{l'})/(x_l-x_{l'})$ does not contribute to the uniform current. Then, the meaning of Eq. (\ref{einstein}) 
becomes even more evident,
\begin{eqnarray}
I(t) & \simeq &  \sum_{l,l'}   \frac{  (x_l-x_{l'})^2}{2}  \Gamma^0_{l,l'}  \left[   -  \frac{n_l-n_{l'}}{x_l-x_{l'}} + \frac{n_l+n_{l'}}{2}  \beta  F \right] . \nonumber \\
\label{einstein1}
\end{eqnarray}
We may further simplify the analysis by neglecting the explicit
momentum dependence of $ |\eta_{l,l',q} |^2 $,  
\begin{equation}
|\eta_{l,l',q} |^2 \simeq |\eta_{l,l'} |^2 = \frac{1}{L} \sum_q |\eta_{l,l',q} |^2= g^2 \sum_i  ( \phi_{l'i}  \phi_{li})^2.
 \label{appro}
\end{equation}
and assume an uniform bosonic density of states  
$1/L \sum_q \delta(\omega-\omega_q)  \simeq  \frac{1}{\Omega} \theta(\Omega-\omega) $  where $\Omega =\omega_0+2t_B$.
Then the transition rates at $\beta \rightarrow 0$ read
\begin{equation}
\Gamma^0_{l,l'} \simeq   \pi  |\eta_{l,l'} |^2  \frac{1}{\Omega}\theta(\Omega-|\Delta_{l,l'}|).
\label{gamma}
\end{equation}

In order to explain the interplay between strong disorder and strong driving we calculate
$\Gamma_l=\sum_{l'\ne l}   \Gamma^0_{l,l'}$ representing the inverse lifetime of a particle occupying the Anderson 
state $| l \rangle$, $\tau_l=1/\Gamma_l $.  As stressed before, it is  essential to avoid averaging of $\Gamma_l$ over disorder. 
Instead one may consider it as a random variable and discuss the probability density $f_{\Gamma}(\Gamma_l)$ or equivalently $f_{\tau}(\tau_l)$. 
Without driving \cite{prelovsek2018a,kozarzewski18}, the qualitative transport properties can be read out from the latter distribution 
using the random--trap model \cite{bouchaud89}. Namely, for $f_{\tau}(\tau_l)\propto 1/\tau_l^{\alpha+1}$ the normal diffusive 
transport exists for $\alpha \ge 1$, whereas for $0<\alpha < 1$ the particle spreads subdiffusively with $\langle x^2(t) \rangle \propto t^{\gamma}$
and $\gamma=2 \alpha/(1+\alpha)$. By  comparing the cumulative distribution functions one finds that  the distribution 
$f_{\tau}(\tau_l)\propto 1/(\tau_l)^{\alpha+1}$  corresponds
to $I(\Gamma)=\int_0^{\Gamma} f_{\Gamma}(\Gamma_l) {\rm d} \Gamma_l \propto \Gamma^\alpha$, so the type of dynamics  (i.e., the value of $\alpha$) can be recognized directly from $I(\Gamma)$.

\begin{figure} 
 \includegraphics[width=\columnwidth]{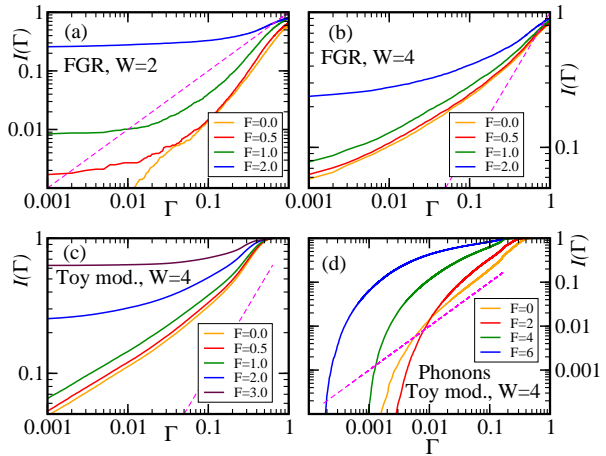}
\caption{ Cumulative distributions $I(\Gamma)$ in  panels a) and b)  for different $F$ were obtained  from Eqs.~(\ref{gam1})-(\ref{gam2}),
based on FGR.  In panel (c) we show  results for the simplified (toy model)
rates as given by, Eq. (\ref{gamma}).  The dashed lines in all panels represent the threshold for 
normal diffusion,  $I(\Gamma) \propto \Gamma$.}   
\label{fig3}
\end{figure}

Fig.~\ref{fig3} shows the FGR results for $I(\Gamma)$. Results in panels a) and b) are obtained from 
Eqs.~(\ref{gam1})-(\ref{gam2}), whereas panel c) shows results for the simplified transition rates, Eq.~(\ref{gamma}), labeled as {\it Toy-model}.  The dashed lines show $I(\Gamma) \propto \Gamma$ hence they mark the threshold for the normal diffusive transport.
Comparing panels b) and c)  it becomes quite evident that simplification via Eq.~(\ref{appro}) is very accurate. One may also see 
that upon increasing $F$, the exponent $\alpha$  decreases and eventually becomes vanishingly small for $F \geq 2$.  The latter result means 
that $f_{\Gamma}(\Gamma_l)$ acquires a $\delta(\Gamma_l)$ contribution or, in other words, that some states remain localized despite being 
coupled to HCB.  Since this effect originates from strong electric field, it is legitimate  to attribute it  as the  Stark localization.

Finally, we argue that the Stark localization doesn't occur if the particle is coupled to standard bosons, e.g., phonons.  
Then, the multi--boson processes significantly contribute to the transition rates, whereas previously,  such contributions were 
strongly suppressed by the hard core--effects.  As argued for  the standard bosons  \cite{prelovsek2018a}, the rigid cut-off in Eq.~(\ref{gamma}) 
should be replaced by a smooth exponential cut-off, $\theta(\Omega-|\Delta_{l,l'}|) \rightarrow \exp(-|\Delta_{l,l'}|/\Omega)$. 
This seemingly harmless modification, changes the distribution of the transition rates, as shown in Fig. \ref{fig3}d. Even for very strong 
drivings and for very strong disorder the transitions rates are bounded from below $\Gamma_l \ge \Gamma_{\rm min}$. Therefore, 
the diffusion constant might be very small but nevertheless nonzero. Although the subdiffusive transport may show up for a quite 
long time-window,  it is a transient phenomenon that  will eventually be replaced by a normal diffusion, irrespectively of $W$ or  $F$.   

The GER is expected to hold true within the LR theory \cite{robin2009,bouchaud89}, whereas we have demonstrated that it remains applicable (at least qualitatively) for stronger drivings when assumptions of LR are not fulfilled in an obvious way. We note that in deriving Eq.~(\ref{einstein}) from the master equation (\ref{req}) we have not assumed that driving is weak or that the transport is normal diffusive, since the anomalous transport  and nonlinear (in $F$) effects are encoded in $\Gamma^0_{l,l'}$, see Eq. (\ref{gamma}). 
The essential step in our derivation consist in expanding $\tanh[\frac{ \beta (\epsilon_l-\epsilon_{l'})}{2}]$ linearly in $F$. Here, $\beta$ is the temperature of the bosonic bath so the bosonic bath must be in  equilibrium (or close to equilibrium) and the
expansion is justified when $\beta $ is small.

One may also expect that the present model is oversimplified in that it describes only single quantum particle coupled to mutually noninteracting HCB. A more realistic description should account for nonzero density of fermions which, in turn, may induce effective interactions among the HCB. In order to check how the latter phenomena affects results presented in this manuscript, we have studied a similar system but with  a boson-boson interactions. Results shown in the Supplemental Material  \cite{supp} demonstrate 
that our qualitative conclusions hold true also for mutually interacting HCB \cite{bonca2018}.

{\it Conclusions. --}
We have studied the transport properties of a quantum particle  that propagates along a disordered chain and is coupled to hard--core bosons,
whereby such a case can be considered as the simulation of the charge motion in the spin background in strongly correlated disordered 
systems. Without external driving, the particle exhibits anomalous subdiffusive propagation with vanishing diffusion constant. 
The main goal was to  establish whether a generalized Einstein relation holds true in such a system. Namely,  we have studied the 
relation between the drift originating from the electric field ($F$) and the spread of the particle density which grows in time mainly due to its 
inhomogeneous spatial distribution. Both quantities were determined in the presence of $F$. The GER was  shown to hold true in the LR regime (as
expected) but also within the quasiequilibrium  evolution,   when the energy (temperature) increases in time due to driving. 
In particular, we have demonstrated that a single exponent $\gamma$ characterizes the anomalous dynamics of the spread
$\langle x^2(t) \rangle_F $ and the drift $\langle x(t) \rangle_F \propto \langle x^2(t) \rangle_0  \propto  t^{\gamma}$. Quite unexpectedly, 
the latter similarity between  $\langle x^2(t) \rangle_F$ and $\langle x(t) \rangle_F$ holds true also for much stronger fields beyond the regime 
of quasiequilibrium  evolution. However for strong fields, $\gamma$ decreases with increasing $F$ and eventually, for very strong $F$,  it vanishes 
marking the field-induced Stark localization.  Qualitatively, all these properties were demonstrated to follow from the distribution of the transition rates between the Anderson states.  
 We have also argued that the subdiffusive transport and  localization should not occur when  the particle is coupled to 
regular itinerant  bosons with unbounded energy spectrum. In the latter case,
strong driving may lead to a transient slowing down of the particle dynamics, nevertheless the asymptotic transport is expected to be normal diffusive. We cannot exclude that within a more accurate treatment of the multi-boson processes the normal diffusion eventually shows up also in the HCB model, however, for much longer times than for regular bosons.

\acknowledgments  
 P.P. and J.B. acknowledge the support by the program P1-0044 of the
Slovenian Research Agency. M.M. is supported by the National Science Centre, Poland via project 2016/23/B/ST3/00647. J.B. acknowledges support from CINT. This work was performed, in part, at the Center for Integrated Nanotechnologies, a U.S. Department of Energy, Office of Basic Energy Sciences user facility.

\bibliography{ref_mbl}

\newpage
\ 
\newpage

\setcounter{equation}{0}
\setcounter{figure}{0}
\setcounter{table}{0}
\makeatletter
\renewcommand{\theequation}{S\arabic{equation}}
\renewcommand{\thefigure}{S\arabic{figure}}

\centerline{{\bf \large Supplemental Material\\}}

In the Supplemental Material  we discuss the finite--size (FS) effects for the exact diagonalization (ED) results. Further on, we compare numerical results obtained from ED and the rate equations  (RE) where transition rates are estimated from the 
Fermi golden rule (FGR).  Finally, we present results for a chain where the hard-core boson (HCB) mutually interact with each other.

\section{Details of numerical calculations and  finite--size effects}  
\label{FS}
Numerical ED calculations were performed using full Hilbert spaces on one-dimensional clusters with periodic boundary conditions. First, we have used standard Lanczos procedure to compute ground state wavefunction $\psi(t=0)$  for each realization of random potentials $h_j\in [-W,W]$. 
The time evolution of the initial state was then calculated by step--vise change of the Peierls phase $\delta \phi(t)=F\delta  t$ in small time increments $\delta t = 0.02$ 
at each step using  Lanczos basis generating the time--evolution $\psi (t - \delta t) \to \psi(t)$. Special attention was used to check the independence  of results against the choice of the time--step. The dependence of results on system  sizes is presented in  Fig.~\ref{figS1} where we show $R(t)$ using different system sizes ranging from $L=10$ to $L=15$. Curves, representing different system sizes are nearly overlapping, indicating 
that finite-size effects  are inessential at least for qualitative conclusions. In addition,  we display  in Fig.~\ref{figS3}  results for two different system sizes also for the generalized model with interacting HCB's. 

In  Fig.~\ref{figS1}c we show results obtained from FGR and RE on a set of different system sizes ranging from $L=50$ to $L=400$. Here, we average in fixed random-configuration results over all 
initial sites  in the system (away from boundaries) and in addition over  $N_s \sim 10$ random samples. Results thus become $L$ independent provided that  $\langle  x^2(t) \rangle \ll L^2 $.

\section{Direct comparison between exact-diagonalization and the Fermi-golden-rule results}

\begin{figure} 
 \includegraphics[width=\columnwidth]{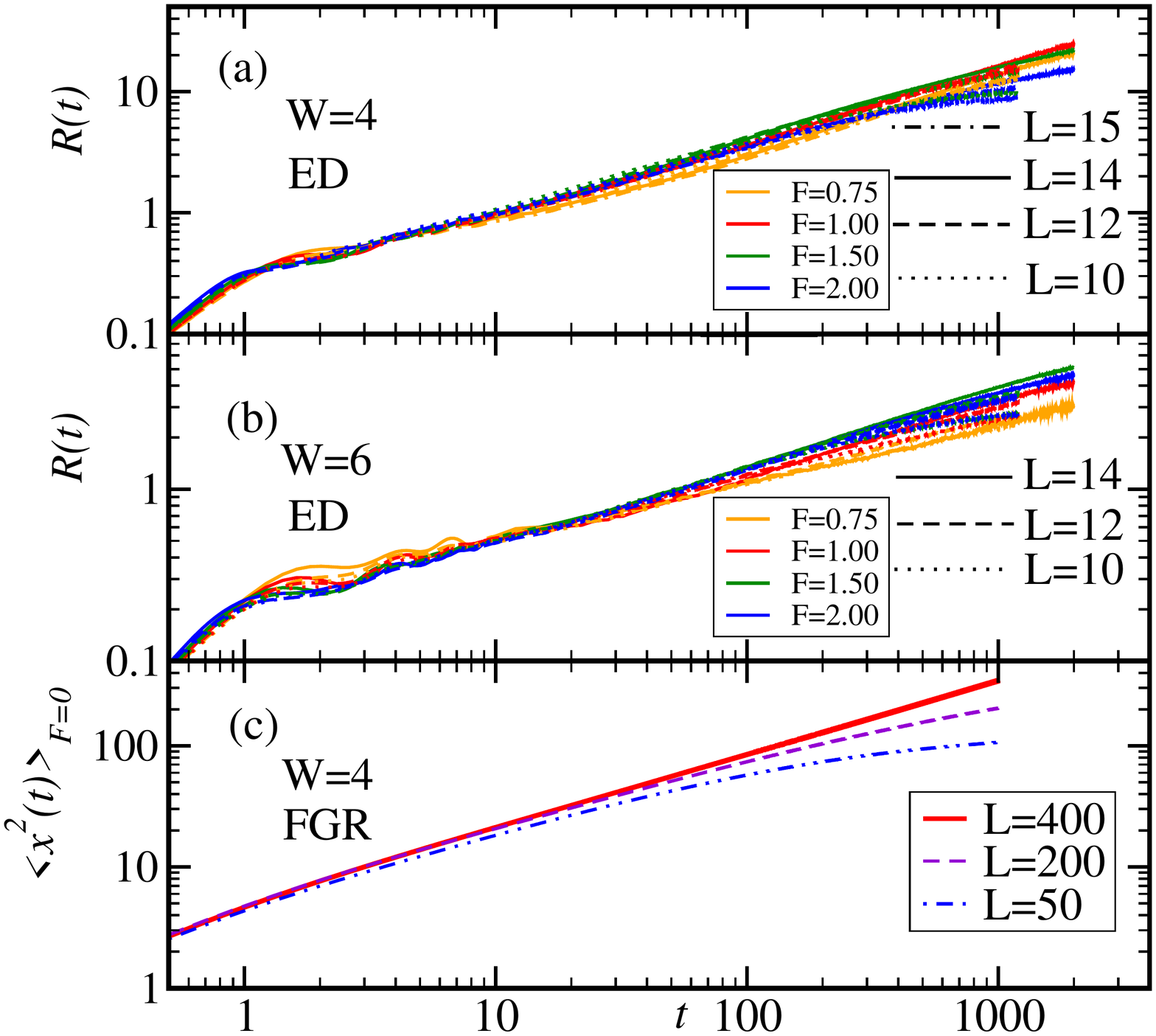}
\caption{ $R(t)$ using ED  for system sizes $L=10, 12, 14$ in (a) and (b) and also $L=15$ in (a). Results are shown for identical parameters of the model as in Fig.~1 of the main paper: $\omega_0=g=1$, and $t_g=0.5$. In (a) there are 4  nearly overlapping curves for each value of $F$ representing results for $L=12,14,$ and 15.  Similarly  in (b),  there are 3 nearly overlapping curves for each $F$.  In (c) we present RE data based on FGR for  $<x^2(t)>_{F=0}$ for different system sizes ranging from $L=50$ through 400.} 
\label{figS1}
\end{figure}

In this section we  discuss numerical results which enable quantitative comparison of ED and RE [Eq. (7) in the main text] with transition rates obtained from the FRG [Eq. (5) in the main text].
Although  the RE may be formally applied to a particle propagating on a finite lattice, this particle must be coupled to an infinite bosonic bath, since
the FGR requires a continuous density of states.
In contrast to ED, our system sizes $L\sim 400 $ are free from the FS effects, as long as the spread  $\langle  x^2(t) \rangle \ll L^2 $.
In the preceding section we have shown that numerical result for ED weakly depend on $L$. Therefore, we compare numerical results for the largest system sizes, 
i.e. ED for $L=14$ with  RE for $L=400$, see  Fig.~\ref{figS1}(c).

\begin{figure} 
 \includegraphics[width=\columnwidth]{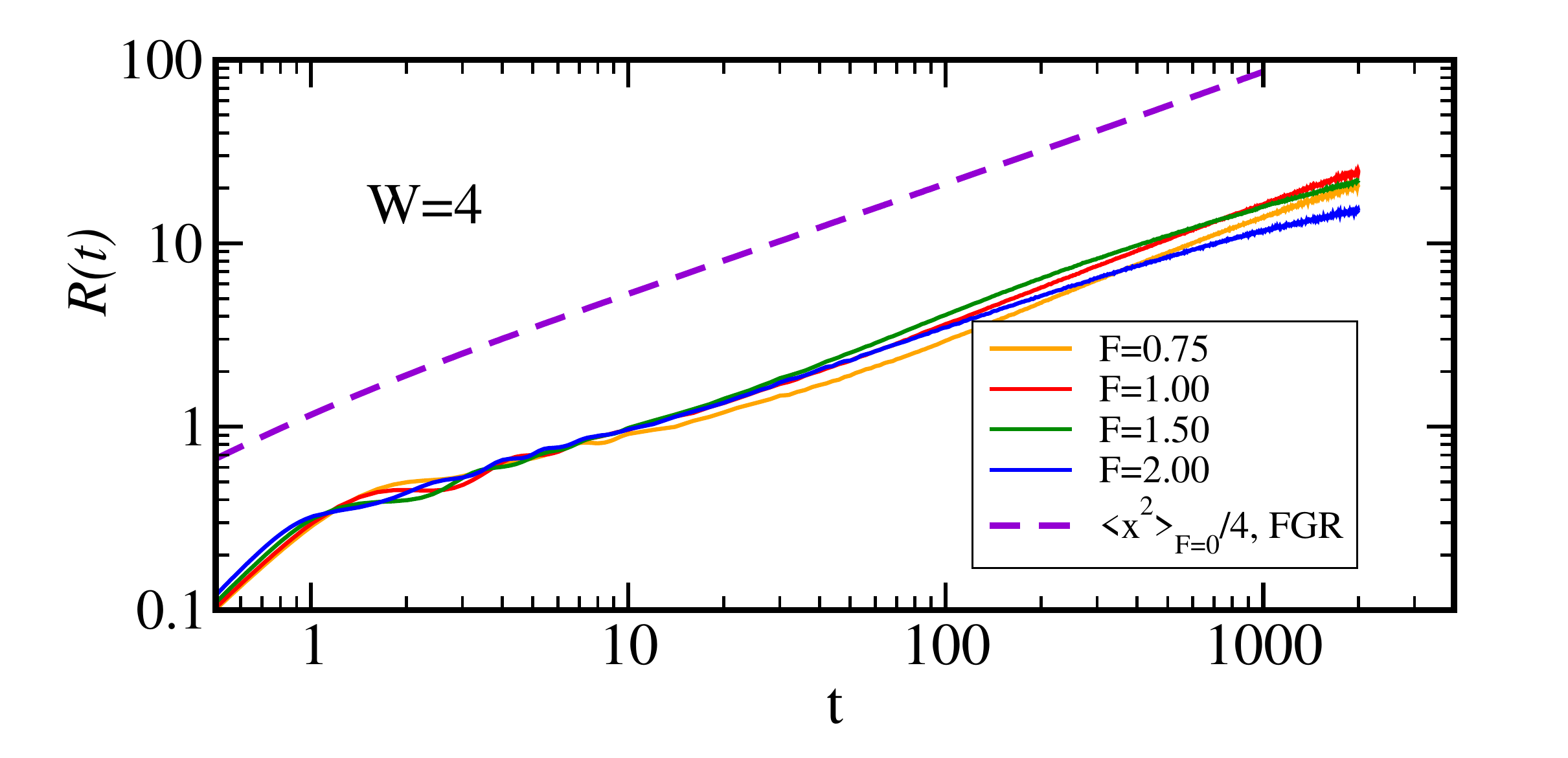}
\caption{ The same as  Fig. \ref{figS1}a but compared with   $\frac{1}{4}  \langle x^2(t) \rangle _{F=0}$    (violet dashed  line) as obtained from RE based on FGR  without  driving, $F=0$. }   
\label{figS2}
\end{figure}

\begin{figure} 
 \includegraphics[width=\columnwidth]{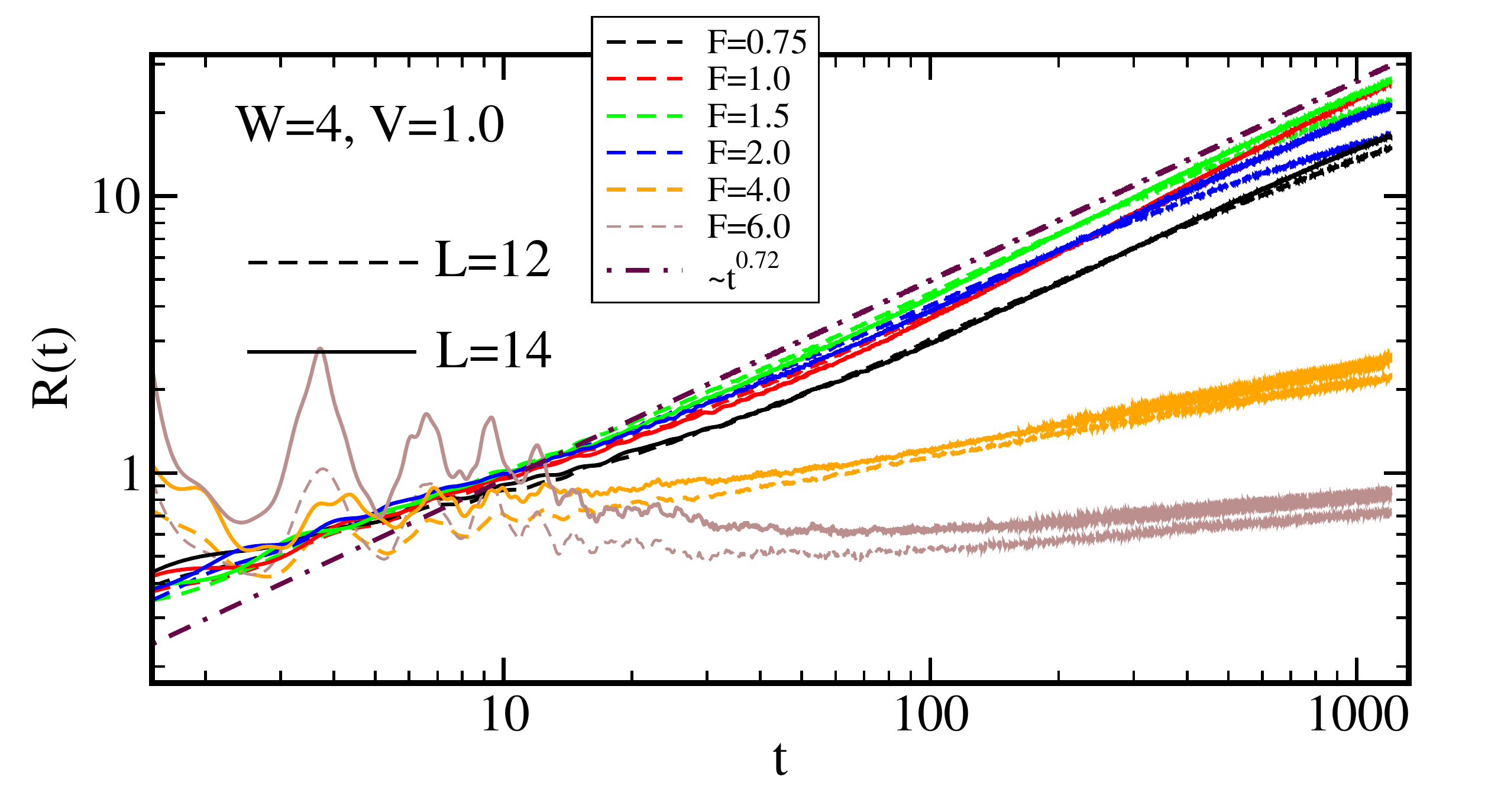}
\caption{$R(t)$ for $W=4$ and $V=1$ for different $F > 0$, up to maximal $F=6$ using ED on $L=12$ (dashed lines)  and 14 sites (full lines). Dot-dashed line represents $R(t)\propto t^{0.72}$. Other parameters of the model are the same as in Fig.~\ref{figS1}.}   
\label{figS3}
\end{figure}

Within the high-temperature expansion one obtains the kinetic energy of a single particle, $E_k=-2 \beta$.  Introducing this relation into  Eqs. (1) and (3)  in the main text one obtains the Einstein relation
$ R(t)=\frac{1}{4} \langle x^2(t) \rangle  $, where  $R(t)=- \Delta  E(t)/ F^2  E_k(t)$.  Figure \ref{figS2} shows $R(t)$ obtained from ED  in comparison with the spread  $\frac{1}{4} \langle x^2(t) \rangle_{F=0}$ 
obtained from RE together with FGR.  The quantitative comparison might seem dissatisfactory, since the results clearly differ.  However, we notice that differences between both methods
arise during the short time dynamics, $t   \lessapprox 5$, whereas for  longer times both methods give consistent results since curves remain parallel to each other. 
It is rather clear that the latter differences cannot originate from the FS effects, since FS should be most visible at longer times when the spread of particle becomes comparable to 
the system size. The differences between both methods may partially originate from different initial states: in the RE we start from a chosen Anderson state while
in ED from the ground state of $H$, Eq.~1 (in the main text), before the onset of driving.  Nevertheless, the origin of these difference should be attributed predominantly to the failure of the RE for short--times.  Since RE represent the classical master
equation it must fail if the propagation time is smaller than the decoherence time. To conclude this section, we notice that the RE and ED results obtained for the largest accessible system sizes are consistent except for the short time-window, where the failure of the RE is well justified and expected.

\section{Interacting Hard--Core Bosons} 
\label{FS}
As argued in the main text, in the case of  nonvanishing density of fermions, the fermion-boson interaction  may induce an effective interactions among the HCB.
Following this arguments, in this section we present additional test for the robustness of the GER by introducing interaction $V$ between HCB's located on neighboring sites.  
The introduction of  small  $V$  lifts the degeneracy among itinerant  HCB states.  Further increase of $V$  leads to  slowing down 
the propagation of excitations while it  concurrently increases the energy of excitations  in the HCB subspace.   Generalization of the original model to interacting HCB's thus represents a suitable  model to test the stability of GER. 

In Fig.~\ref{figS3} we present  results for  $R(t)$ obtained for a generalized interacting HCB model:
\begin{eqnarray}
H_U &=& - t_h \sum_j (c^\dagger_{j+1}  c_j + {\mathrm h.c.}) + \sum_j h_j n_j  + g \sum_j n_j 
( a^\dagger_j + a_j ) \nonumber \\
&+& \omega_0 \sum_j a_j^\dagger a_j - t_b  \sum_j ( a_{j+1}^\dagger a_j + \mathrm{H.c.}) \nonumber \\
&+& V\sum_j a_j^\dagger a_j a_{j+1}^\dagger a_{j+1}. 
\label{hams} 
\end{eqnarray}   
Apart for larger extracted $\gamma=0.72$, in comparison to $\gamma=0.58$ in the case of itinerant HCB's, results are very similar to the ones, presented in Fig.~2(a) of the main text. For small driving up to  $ F\lesssim 2.0$ we observe scaling, characteristic for GER. The increase of $\gamma$ in comparison to itinerant HCB's may be attributed to a combination of lifting of the degeneracy as well as to the increased energy of HCB excitations that in turn increases the ability of the HCB subspace to absorb energy from the driving.  For  stronger driving $F>2$, exponent $\gamma$ (as well as particle dynamics) is significantly reduced
and eventually, for very strong $F \sim 6 $ the particle becomes almost localized.

\end{document}